\documentclass[12pt]{article}
\pdfoutput=1
\usepackage{epsfig}
\usepackage{amsfonts}
\usepackage{amssymb}
\usepackage{xcolor}
\usepackage{amsmath,amsfonts,amssymb,amsthm,mathrsfs,mathtools} 
\graphicspath{{pictures/}}

\begin{document}
\newcommand{\nc}{\newcommand}
\nc{\beq}{\begin{equation}} \nc{\eeq}{\end{equation}}
\nc{\beqa}{\begin{eqnarray}} \nc{\eeqa}{\end{eqnarray}}
\nc{\R}{{\cal R}}
\nc{\A}{{\cal A}}
\nc{\K}{{\cal K}}
\nc{\B}{{\cal B}}
\nc{\al}{\alpha}
\nc{\bt}{\beta}
\nc{\gm}{\gamma}
\nc{\alh}{ \hat{\alpha} }
\nc{\yh}{ \hat{y} }
\begin{center}

{\bf \Large  New Solutions of RG Equations  \\[0.3cm] for  $\alpha_s$ and $y_{top}$} \vspace{1.0cm}

{\bf \large A.S. Fedoruk$^1$ and D. I. Kazakov$^2$} \vspace{0.5cm}

{\it $^1$Moscow Institute for Physics and Technology, Dolgoprudny, Russia, \\
$^2$Bogoliubov Laboratory of Theoretical Physics, Joint
Institute for Nuclear Research, Dubna, Russia} 

\vspace{0.5cm}

\abstract{We construct simple analytical solutions of the RG equations for the running couplings $\alpha_s$ and $y_{top}$  in the asymptotic regime. These solutions have an explicit form, contain only logarithms and no special functions, and subsequently sum up the leading, subleading, etc logarithms  in all orders of PT.  While the effect of $y_{top}$ on the running of $\alpha_s$ happens to be negligible, the role of $\alpha_s$ on the running of $y_{top}$ is essential, the account of higher orders gives a noticeable  contribution.
}
\end{center}

Keywords: RG equations, Running coupling,  Leading Log approximation, Next-to Leading Log Approximation, Higher orders of PT

 \section{Introduction}
 
 In a recent paper~\cite{IKT} we proposed a strategy that allows one to sum up the leading, NL, NNL, etc logs for the running coupling, resulting in  a relatively simple scheme that provides an explicit analytical form at each stage. We demonstrated how this strategy works in a theory with a single coupling like QCD. In this note, we extend the proposed approach to the case of two couplings, namely we consider the real case of the Standard Model with the strong and top Yukawa couplings. All the other couplings  are relatively small and we ignore them in our analysis. 
 
 Remind the main issues of the advocated approach. Our main point is not to solve the RG equations written in a given order of PT exactly but first to linearise them and then solve the
 linear equations in subsequent orders. These new explicit solutions provide a summation of the  corresponding logs in the Leading, Nex-to-Leading, etc approximation. In a theory with a single coupling, this results in  a regular expansion containing only elementary logs and no other special functions. Moreover, further summation and improvement are possible, leading to the same logs but with more complex arguments~\cite{IKT}. 
 It should be stressed  that the usual approach to  solving the non-linear RG equations in higher orders does not allow for analytical solutions and sums not only the corresponding log terms but also  includes some amount of subleading logs (see e.g. Appendix in \cite{BMS}). The advantage of our approach is that at each stage we sum just the needed logs and do it analytically.
  
 The case of two couplings happens to be more complicated since the one-loop RG equation for the Yukawa coupling has a tricky solution even though it contains only logarithms. As a result, the solutions of our linearised  equations in higher orders contain, in addition to the logarithmic part, a relatively small addendum described by a special function. In what follows, we ignore this part in the asymptotic regime due to its numerical suppression. A relatively simple analytic form allows for  further summation of the leading terms, just like in the one-coupling case, which results in more stable asymptotic solutions.
 
\section{Perturbative Expansion and New Solutions}
  
  Consider the Standard Model where only the strong gauge and top Yukawa interaction are  taken into account and all the other couplings are ignored. Then  the RG equations for the running couplings  have the form
  \beqa \label{eq:al}
    \frac{d\al(L)}{dL} &=& \beta_0\al^2 + \al^2 \left( \bt_{10} \al + \bt_{01} y \right) + \al^2 \left( \bt_{20} \al^2 + \bt_{11} \al y + \bt_{02} y^2 \right) + \ldots ,\\
    \frac{dy(L)}{dL} &=&y \left( \gm_{10} \al+ \gm_{01} y \right)  + y \left( \gm_{20} \al^2 + \gm_{11} \al y + \gm_{02} y^2 \right) \nonumber \\
    && \hspace{2.8cm}+ y \left( \gm_{30} \al^3 + \gm_{21} \al^2 y + \gm_{12}\al y^2+\gm_{03}y^3\right)+\ldots,
    \label{eq:y}
\eeqa
with the boundary conditions $ \al(0) = \al_0, \ y(0) = y_0,$
 where we use the notation\\ $\alpha=g_s^2/16\pi^2,\, y=y_t^2/16\pi^2,\, L=\ln(Q^2/\mu^2)$.

In the SM with 6 flavours, the coefficients  in the above equations up to three loops in the $\overline{MS}$ scheme take the values~\cite{Bednyakov2021,RGBeta}
\beqa \label{coeff1}
  \beta_0 &= &-7,\, \gm_{10} = -8,\, \gm_{01} = 9/2, \, \bt_{10} = -26,\, \bt_{01} = -2, \nonumber \\
 \gm_{20} &=& -108,\, \gm_{11} = 36,\, \gm_{02} = -12,\, \bt_{20} = 65/2, \,
    \bt_{11} = -40,\, \bt_{02} = 15,\\ 
    \gm_{30} &=& -4166/3 + 640 \zeta(3),\, \gm_{21} = 3827/6 - 228 \zeta(3), \,  \gm_{12} = -157,\,
  \gm_{03} = 339/8. \nonumber 
  \eeqa
  
 Following our approach, we look for solutions  of RG equations (\ref{eq:al}, \ref{eq:y}) in the form of a loop expansion
 \beqa
  \al &= &\sum_{\ell=1}^{\infty} \al_\ell = \al_1 + \al_2 + \ldots  \nonumber \\
  y &= &\sum_{\ell=1}^{\infty} y_\ell = y_1 + y_2 + \ldots \\
  \al_\ell& \sim& y_\ell \;\sim\; \sum_{n=\ell}^{\infty} (\al_0 + y_0)^n L^{n-\ell} \;\sim\; (\al_0 + y_0)^\ell + (\al_0 + y_0)^{\ell+1} L + \ldots \nonumber 
   \eeqa
 with the boundary conditions  that are convenient to choose like 
$ \al_1(0) =\al_0, \ y_1(0) = y_0, \ 
\al_\ell(0) = y_\ell(0) = 0 \ \  (\ell > 1)$.
Here $(\al_0+y_0)^n$ symbolically denotes all possible powers of the form $\al_0^{n-k}y_0^k$.
 
 The functions $\al_\ell$ and $y_\ell$ sum up the leading ($\al_1,y_1$), next-to-leading ($\al_2,y_2$), etc logarithms in all orders of PT. They obey the linearised RG equations  derived from  (\ref{eq:al}, \ref{eq:y}) keeping the terms of the same order of magnitude, namely
 \beqa\label{eq:al_y_expanded_equations}
 &&  \frac{d\al_1}{dL}  = \beta_0 \al_1^2,   \\
  && \frac{dy_1}{dL}  = y_1 \left(\gm_{10} \al_1 + \gm_{01} y_1\right),   \\
 &&  \frac{d\al_2}{dL}  = 2 \beta_0 \al_1 \al_2 + \bt_{10} \al_1^3 + \bt_{01} \al_1^2 y_1 , ,\label{a2} \\
&&   \frac{dy_2}{dL} = \left(\gm_{10} \al_1 + 2 \gm_{01} y_1\right) y_1 + \gm_{10} y_1 \al_2 + \gm_{20} \al_1^2 y_1 + \gm_{11} \al_1 y_1^2 + \gm_{02} y_1^3 ,\label{y2}\  \\
 && \ldots    \nonumber \\
&& \left. \frac{d\al_\ell}{dL} \right|_{\ell>1} =2 \beta_0 \al_1 \al_\ell + \Phi_\ell ,\label{alpha}\\
 &&  \left. \frac{dy_\ell}{dL} \right|_{\ell>1} = \left(\gm_{10} \al_1 + 2 \gm_{01} y_1\right) y_\ell + \gm_{10} y_1 \al_\ell + \Psi_\ell\label{ygrek},
\eeqa
where the functions $\Phi_\ell$ and $\Psi_\ell$ contain only $\al_{\ell^\prime}$ and $y_{\ell^\prime}$ with $\ell^\prime<\ell$. 
Notice that the equations for $\left.\al_\ell \right|_{\ell>1}, \left.y_\ell \right|_{\ell>1}$ turned out to be linear and allow an explicit integral form for the  solution.

\subsection{$\ell=1$}

\begin{figure}[ht]
  \centering
  \includegraphics[width=0.6\textwidth]{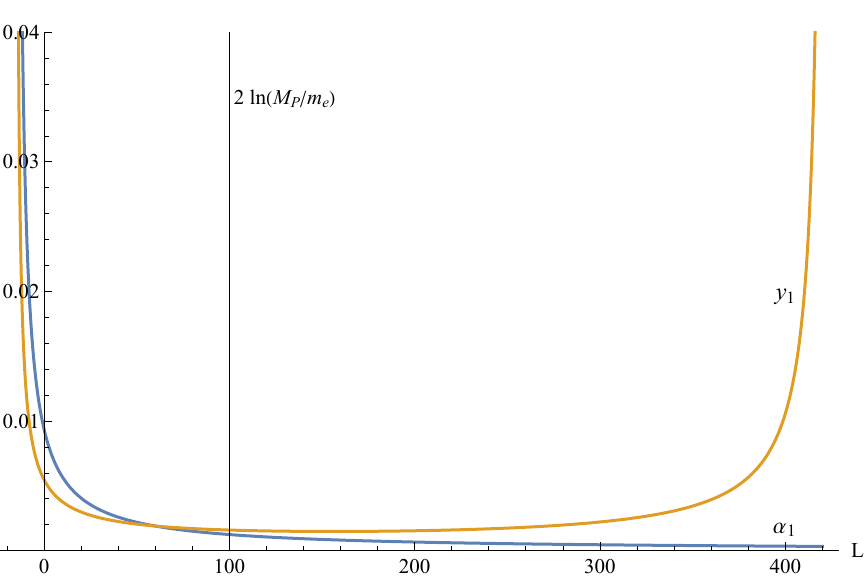}
  \caption{The running of $\al_1,y_1$ in the SM}\label{fig:al_y_1}
\end{figure}

The equations for $\al_1,y_1$ are ordinary one-loop RG ones and have the following  well-known solutions~\cite{Ibanez1985, KazakovYukawa}:
\begin{equation}\label{eq:al_y_1_solutions}
  \left\{\begin{array}{l}
    \al_1 = \dfrac{\al_0}{1-\beta_0\al_0L} \\[10pt]
    y_1 = \dfrac{(\gm_{10}-\beta_0)\al_0 y_0}{-\gm_{01} y_0 \left(1-\beta_0\al_0L\right) + (\al_0 \left(\gm_{10}-\beta_0\right) + \gm_{01} y_0) \left(1-\beta_0\al_0L\right)^{\scalebox{0.8}{$\nu$}} }\,, \quad \nu = \gm_{10}/\beta_0.
  \end{array}\right.
\end{equation}
The corresponding plot for $\al_0= 9.39\cdot 10^{-3},\, y_0 = 5.48\cdot 10^{-3},\, \mu=M_Z$~\cite{PDG}  is shown in Fig.\ref{fig:al_y_1}. 

As one can see, there is a Landau pole for the Yukawa coupling at $L=1/(\beta_0\al_0)\approx -15.2$ and  an additional one  at $L\approx 420$, which, however, lies much further than the Planck scale. 

\subsection{$\ell > 1$}

Since equations (\ref{alpha},\ref{ygrek}) for $\ell > 1$ are linear, one can write down the general solutions keeping in mind the initial conditions $\al_\ell(0) = y_\ell(0) = 0$:
\beqa
 \displaystyle\al_\ell &=&  \al_1^2 \int_{0}^{L} \frac{\Phi_\ell}{\al_1^2} \:dL^\prime, \\
  \displaystyle y_\ell &=&  \frac{y_1^2}{\al_1^\nu} \int_{0}^{L} \frac{\al_1^\nu}{y_1^2} \left(\gm_{10} y_1 \al_\ell + \Psi_\ell\right) \:dL^\prime,
\eeqa
where for the evaluation of some integrals the following relation was used:
\beq\label{eq:1_over_y1}
  \dfrac1{y_1} = \dfrac{-\gm_{01}}{\gm_{10}-\beta_0} \dfrac1{\al_1} + \dfrac{\al_0(\gm_{01}-\beta_0)+y_0\gm_{01}}{\al_0y_0(\gm_{10}-\beta_0)}\left(\dfrac{\al_0}{\al_1}\right)^\nu.
\eeq

One can check that these solutions indeed reproduce the expansion $\sim \sum (\al_0 + y_0)^n L^{n-\ell}$. However, due to a complicated form of $y_1$ these integrals cannot be evaluated analytically  and one is bounded to take the approximations. We analysed this problem in the lowest orders and found some general pattern.

\subsection{$\ell=2$}

\begin{figure}[ht]
 \centering
 \includegraphics[width=0.7\textwidth]{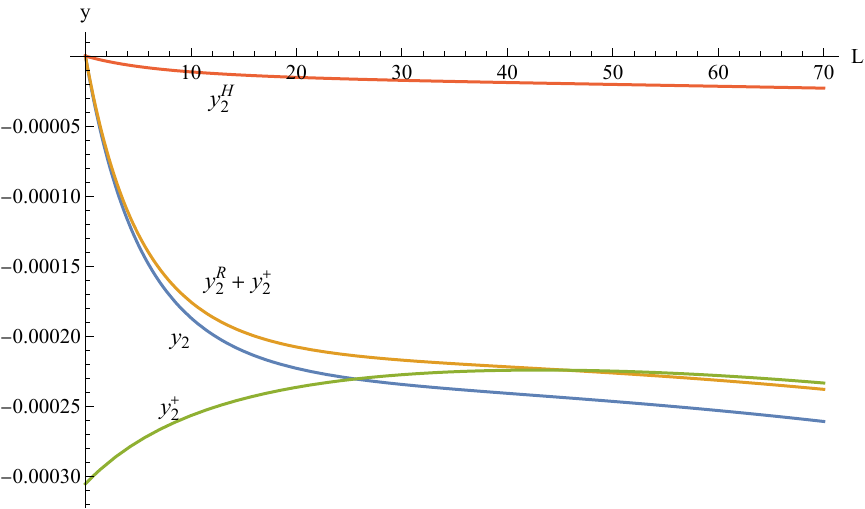}
 \caption{The running of $y_2$. Here different contributions to $y_2$ are shown.\label{fig:y2R+1}}
\end{figure}

Solving equations (\ref{a2},\ref{y2}) for $\al_2$ and $y_2$, we get the following result:
\beqa \label{eq:al_y_2_exact}
  \al_2&=&\al_2^R=\beta_0\al_1^2\left(c_1 \ln(\al_1/\al_0)+c_2\ln(y_1/y_0)\right), \\
  y_2&=&y_2^R+y_2^++y_2^H, \\
  y_2^R&=&y_1\left(\gm_{10}\al_1+\gm_{01}y_1\right)\left(c_1 \ln(\al_1/\al_0)+c_2\ln(y_1/y_0)\right) + y_1(c_3\al_1+c_4y_1),  \nonumber\\
  y_2^+&=&y_1^2\left(\frac{\al_0}{\al_1}\right)^\nu \frac{c_3 \al_0+c_4y_0}{-y_0}, \nonumber \\
  y_2^H&=&c_5 \frac{y_1^2}{\al_1^\nu}\int_{0}^{L}y_1\al_1^\nu dL^\prime.\nonumber
\eeqa
Here $c_i$ are numbers (a combination of coefficients $\beta$ and $\gamma$)(see App.A), $y_2^R$ is a \emph{regular} part of $y_2$, namely one of the solutions of the equation
\beq
\dot y_2^R = (2 \gm_{01} y_1 + \gm_{10} \al_1) y_2^R + \gm_{10} y_1 \al_2 + \gm_{20} y_1 \al_1^2 + \gm_{11} y_1^2\al_1 + (\gm_{01}\bt_{01}/\beta_0)y_1^3,
\eeq
$y_2^+$ is the solution of the \textit{homogeneous} equation
\beq
\dot y_2^+ = (2 \gm_{01} y_1 + \gm_{10} \al_1) y_2^+,
\eeq
such that $y_2^R(0)+y_2^+(0)=0$, and $y_2^H$ is unfortunately neither an elementary function of $\al_1,y_1$, nor an elementary function of $L$ but rather some \emph{hypergeometric} expression. Fortunately,  $y_2^H$ does not contribute much to the running of $y_2$, as can be seen from Fig.\ref{fig:y2R+1} and we will ignore it in what follows.

\subsection{$\ell=3$}

By analogy with the case of $\ell=2$, the corresponding solutions of the linearised equations for $\ell=3$ take the form
\beqa
  \al_3&=&\al_3^R+\al_3^++\al_3^H, \\
  \al_3^R&=&\beta_0^2\al_1^3(c_1\ln(\frac{\al_1}{\al_0}) + c_2\ln(\frac{y_1}{y_0}))^2 + \al_1^2 \left(\bt_{10} \al_1 + \bt_{01} y_1\right)(c_1 \ln(\frac{\al_1}{\al_0})+c_2\ln(\frac{y_1}{y_0})) \nonumber\\ 
  &+ &\beta_0\al_1 (c_6 \al_1^2 + c_7 \al_1y_1 + c_8 y_1^2 ), \nonumber\\
  y_3&=&y_3^R+y_3^++y_3^H, \\
  y_3^R&=&y_1 (\gm_{01} y_1^2 + \dfrac32\gm_{10} y_1\al_1 + \dfrac{\gm_{10}(\gm_{10}+\beta_0)}{2\gm_{01}}\al_1^2) (c_1 \ln(\frac{\al_1}{\al_0})+c_2\ln(\frac{y_1}{y_0}))^2 \nonumber\\ 
  &+ & y_1 (c_9 y_1^2 + c_{10} y_1\al_1 + c_{11}\al_1^2) (c_1 \ln(\frac{\al_1}{\al_0})+c_2\ln(\frac{y_1}{y_0}))+ y_1 (c_{12}y_1^2 + c_{13}y_1 \al_1 + c_{14} \al_1^2).\nonumber 
  \eeqa
  
Once again, the parts $\al_3^H$ and $y_3^H$ are not expressed via elementary functions and are numerically inessential, so we ignore them in what follows;  $\al_3^+$ and $y_3^+$ are given  by long expressions containing non-integer powers of the coupling (analogously to $y_2^+$ in \eqref{eq:al_y_2_exact}). Being not essential for the demonstration of our method, we do not provide their  exact form.

\section{RG Invariant Parametrization}

It is useful to rewrite the obtained solutions expressing them via RG invariants. In the case of a single coupling theory like QCD, it is the well-known invariant $\Lambda$~\cite{PDG,Deur2016}. In the case of two couplings, one has the other invariant $K$ which characterises the curve in the phase space of  these couplings.   Both parameters vary with the number of loops. 
\subsection{$\ell=1: \hat L_1, K_1$}

The one-loop solutions can be rewritten in terms of scale  invariants as
\beqa 
\label{eq:al_y_reduced_1}
   \hat  \al_1 &=& \dfrac1{-\beta_0\hat L_1}, \\
    \hat y_1 & = &\dfrac{\gm_{10}-\beta_0}{\beta_0\gm_{01}\hat L_1 + K_1(\gm_{10}-\beta_0)\hat L_1^\nu},
  \label{eq:al_y_reduced_11}  
\eeqa
where 
\beqa
\label{eq:al_y_shift_1}
    \hat L_1 &=& L-\dfrac1{\beta_0 \al_0} = \ln(Q^2/\Lambda_1^2), \ \ \ K_1 = \dfrac{\al_0(\gm_{10}-\beta_0)+y_0\gm_{01}}{\al_0y_0(\gm_{10}-\beta_0)}(-\beta_0\al_0)^\nu.
\eeqa

Practical significance of such formulation if that now $\Lambda$ and $K$ \emph{do not depend on the renomalization scale} $\mu$ along the RG curve and actually parametrise the curve itself.

\subsection{$\ell= 2$ and $\ell=3$}

In order to obtain $\hat L$ for higher loop orders, one has to expand the beta-function and use the corresponding RG equations:
\beqa
\label{eq:Lh_123_integral_calculation}
  \hat L-L&=& \ln(\mu^2/\Lambda^2) = \int^{\al_0}\dfrac{d\al}{\beta(\al)},\\
  \hat L_1-L &=& \int^{\al_0}\dfrac{d\al}{\beta_0\al^2} = -\dfrac1{\beta_0\al_0}, \\
  \hat L_2-L &=& \int^{\al_0}\dfrac{d\al}{\beta_0\al^2+\al^2(\bt_{10}\al+\bt_{01}y)}
  \approx \int^{\al_0}\dfrac{d\al}{\beta_0\al^2}(1- \frac{\bt_{10}}{\beta_0}\al+\frac{\bt_{01}}{\beta_0}y)\nonumber\\
  &&= -\dfrac1{\beta_0\al_0}-(c_1\ln\al_0+c_2\ln y_0),
  \eeqa
  where we have used the one loop equation for $y$: $ \dfrac{y}{\al^2}=\dfrac d{d\al}(\dfrac{\beta_0}{\gm_{01}}\ln y-\dfrac{\gm_{10}}{\gm_{01}}\ln\al)$.
  
  Analogously, for $\ell=3$
\beq
 \hat L_3-L = -\dfrac1{\beta_0\al_0}-(c_1 \ln\al_0+c_2\ln y_0-(c_5\al_0+c_6 y_0 + c_7\dfrac{y_0^2}{\al_0}).
\eeq

The value of $\Lambda$ does not vary  much with the loop order. Taking the values of the parameters from
(\ref{coeff1}), we get the value of $\hat L$ for $\ell=1,2,3$:
$$
\hat L-L =  \ln(\mu^2/\Lambda^2)=\{15.21, 12.73, 12.71\}.
$$
As expected, $\Lambda$ approaches some fixed value with more loops taken into account.

To formulate the algorithm for calculating $K$ for any loop order, we use the procedure similar to  (\ref{eq:Lh_123_integral_calculation}) for the second equation. We start with the equation
\beq
\frac{dy}{dL}-\gamma(\al,y)=0,
\eeq
multiply it by $\dfrac{(-\beta_0\al)^\nu}{y^2}$ and rewrite it as
$$-(-\beta_0\al)^\nu\dfrac{d}{dL}(\dfrac1y) -\dfrac{(-\beta_0\al)^\nu}{y^2}\gamma(\al,y)=-\dfrac{d}{dL}\left(\dfrac{(-\beta_0\al)^\nu}{y}\right) - (-\beta_0\al)^\nu\left(\dfrac{\gamma(\al,y)}{y^2} - \dfrac{\nu \beta(\al,y)}{\al y}\right).
$$

Integrating this relation now over $\int dL$ from $0$ to $L$, we get the integration constant at the boundary
\beqa
K&=& \dfrac{(-\beta_0\al_0)^\nu}{y_0} -(-\beta_0)^\nu \int_{L=0} \left[\dfrac{\al^\nu}{y^2}\gamma(\al,y) - \dfrac{\nu \al^{\nu-1}}{y}\beta(\al,y)\right]dL \nonumber \\
   & = & \dfrac{(-\beta_0\al_0)^\nu}{y_0} + (-\beta_0)^\nu \int^{y_0} \dfrac{\al^\nu}{y^2} dy - (-\beta_0)^\nu\int^{\al_0} \dfrac{\nu \al^{\nu-1}}{y} d\al \label{K}.
\eeqa

Substituting the  one-loop beta- and gamma-functions, we get precisely the one-loop expression \eqref{eq:al_y_shift_1} for $K_1$. One should use the two-loop quantities to calculate the next term $K_2$:
\beqa
 K_2 &=&  \dfrac{(-\beta_0\al_0)^\nu}{y_0} + (-\beta_0)^\nu \int^{L=0} \left[\dfrac{\al^\nu}{y^2} y({\gm_{10}\al}+\gm_{01}y+\gm_{20}\al^2+\gm_{11}\al y+\gm_{02} y^2)\right.\nonumber\\
 &&\hspace{2cm} \left.- \dfrac{\gm_{10}\al^{\nu-1}}{\beta_0 y}\al^2({\beta_0}+\bt_{10} \al+\bt_{01} y)\right]dL 
 = \dfrac{(-\beta_0\al_0)^\nu}{y_0} \nonumber\\
 &+& (-\beta_0)^\nu \int^{L=0}\left[\gm_{01}\al^\nu + (\gm_{20}-\dfrac{\bt_{10}\gm_{10}}{\beta_0})\dfrac{\al^{\nu+2}}y + (\gm_{11}-\dfrac{\bt_{01}\gm_{10}}{\beta_0})\al^{\nu+1} + \gm_{02} y \al^\nu \right]dL\nonumber .
\eeqa

The important step here is to treat the $\gm_{01}\al^\nu$ term inside the integral in the $2$-loop approximation, while the others are in the $1$-loop one, so we get the following contribution to $K$:
$$
\begin{array}{l}
  \displaystyle K_2 = \dfrac{(-\beta_0\al_0)^\nu}{y_0} + (-\beta_0)^\nu \int \left[\dfrac{\gm_{01}}{\beta_0} \al^{\nu-2}\left(\beta_0 \al^2 + \bt_{10} \al^3 + \bt_{01} \al^2 y\right) - \dfrac{\gm_{01}}{\beta_0}\al^\nu\left(\bt_{10} \al + \bt_{01} y\right) +\right. \\[8pt]
   \quad \left. +  \left(\gm_{20}-\dfrac{\beta_{10}\gm_{10}}{\beta_0}\right)\dfrac{\al^{\nu+2}}y + \left(\gm_{11}-\dfrac{\beta_{01}\gm_{10}}{\beta_0}\right)\al^{\nu+1} + \gm_{02} y \al^\nu \right]dL = \\[8pt]
  \displaystyle =  \dfrac{(-\beta_0\al_0)^\nu}{y_0} + \dfrac{\gm_{01}(-\beta_0\al_0)^\nu}{\al_0(\gm_{10}-\beta_0)}  \\[8pt]+ (-\beta_0)^\nu \int \left[ (\gm_{20}-\dfrac{\bt_{10}\gm_{10}}{\beta_0})\dfrac{\al^{\nu+2}}y + (\gm_{11}-\dfrac{\bt_{01}\gm_{10}+\bt_{10}\gm_{01}}{\beta_0})\al^{\nu+1} +
    (\gm_{02} -\dfrac{\gm_{01}\bt_{01}}{\beta_0}) y \al^\nu \right]dL.
\end{array}
$$
To evaluate the integrals, we just plug here $\al_1$ and $y_1$, and also discard the $y \al^\nu$ term for the same hypergeometric reasons as before. Using once more eq.\eqref{eq:1_over_y1}, we arrive at the following expression for $K_2$ :
\begin{equation}\label{eq:K2}
  K_2 = \left(\dfrac{\al_0(\gm_{10}-\beta_0)+y_0\gm_{01}}{\al_0y_0(\gm_{10}-\beta_0)} + \dfrac{c_3\al_0+c_4 y_0}{y_0}\right)(-\beta_0\al_0)^\nu.
\end{equation}

Continuing the same way and taking the three loop beta- and gamma-functions in (\ref{K}), we get for  $K_3$: 
\begin{equation}\label{eq:K3}
\begin{array}{l}
  K_3 = \left(\dfrac{\al_0(\gm_{10}-\beta)+y_0\gm_{01}}{\al_0y_0(\gm_{10}-\beta)} + \dfrac{c_3\al_0+c_4 y_0}{y_0} + \dfrac{\al_0}{y_0}\left(\al_0(c_{13}\!-\!\nu c_5) + y_0(c_{12}\!-\!c_{14})\right)\right)(-\beta\al_0)^\nu. 
\end{array}
\end{equation}
Evaluating the values of $K$ for $\ell=1,2,3$, we find:
$$
K=\{-13.22, -12.76, -12.73\}.
$$
We see that, as expected, $K$ approaches some  fixed value with more loops taken into account.

Now, the two-loop solutions (\ref{eq:al_y_2_exact}) can be rewritten in the invariant form
\beqa
\label{eq:al_y_reduced_2}
    \hat \al_2& =& \beta_0\hat \al_1^2\left(c_1 \ln\hat \al_1+c_2\ln\hat y_1\right), \\
    \hat y_2 &=& \hat y_1\left(\gm_{10}\hat \al_1+\gm_{01}\hat y_1\right)\left(c_1 \ln\hat \al_1+c_2\ln\hat y_1\right) + \hat y_1(c_3\hat \al_1+c_4\hat y_1),\label{eq:al_y_reduced_21}
\eeqa
which differs from $\al_2^R$ and $y_2^R$ by replacement  $$(\al_1,y_1,\ln(\al_1/\al_0),\ln(y_1/y_0))\to(\hat \al_1,\hat y_1, \ln\hat \al_1, \ln\hat y_1).$$

Note that when rewriting the solutions in the form \eqref{eq:al_y_reduced_2}, the initial values of the couplings in the arguments of the logarithms and $y_2^+$ are absorbed into the shifts of $\hat L$ and $K$. Up to a given accuracy one then has
\beqa
  \al_1(L,\al_0,y_0)+\al_2(L,\al_0,y_0)&\approx&\hat \al_1(\hat L_2,K_2)+\hat \al_2(\hat L_2,K_2), \nonumber\\
  y_1(L,\al_0,y_0)+y_2(L,\al_0,y_0)&\approx&\hat y_1(\hat L_2,K_2)+\hat y_2(\hat L_2,K_2).
  \nonumber
\eeqa
In the three-loop order, using the same prescription, one has
\beqa
\label{eq:al_y_reduced_3}
    \hat \al_3& =& \beta_0^2\hat \al_1^3\left(c_1\ln\hat\al_1 + c_2\ln\hat y_1\right)^2 + 
    \hat\al_1^2 \left(\bt_{10} \hat\al_1 + \bt_{01} \hat y_1\right)\left(c_1 \ln\hat \al_1+c_2\ln\hat y_1\right) \nonumber \\
    &+&  
    \beta_0\hat \al_1 \left( c_5 \hat\al_1^2 + c_6\hat \al_1\hat y_1 + c_7 \hat y_1^2 \right),\\ 
     \hat y_3 &= & \hat y_1 \left(\gm_{01} \hat y_1^2 + \dfrac32\gm_{10} \hat y_1\hat \al_1 + \dfrac{\gm_{10}(\gm_{10}+\beta_0)}{2\gm_{01}}\hat \al_1^2\right) \left(c_1 \ln\hat \al_1+c_2\ln\hat y_1\right)^2
     \label{eq:al_y_reduced_31} \\ &+ & \hat y_1 \left(c_8 \hat y_1^2 + c_9 \hat y_1\hat \al_1 + c_{10}\hat \al_1^2\right) \left(c_1 \ln\hat \al_1+c_2\ln\hat y_1\right) +  \hat y_1 \left(c_{11}\hat y_1^2 + c_{12}\hat y_1 \hat\al_1 + c_{13}\hat \al_1^2\right),  \nonumber 
\eeqa

For illustration, we show below the behaviour of  the gauge and Yukawa couplings according to the solutions
(\ref{eq:al_y_reduced_1}-\ref{eq:al_y_reduced_11},\ref{eq:al_y_reduced_2}-\ref{eq:al_y_reduced_21},\ref{eq:al_y_reduced_3}-\ref{eq:al_y_reduced_31}).
\begin{figure}[ht]
 \centering
 \includegraphics[width=0.47\textwidth]{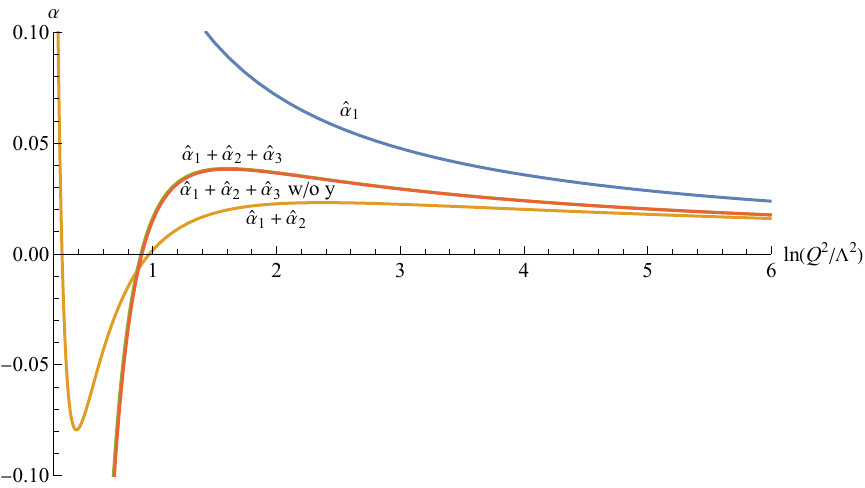}
 \includegraphics[width=0.47\textwidth]{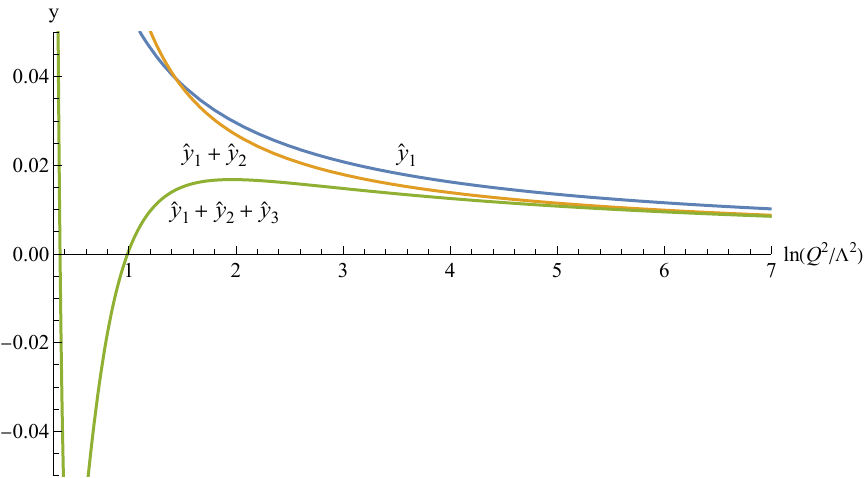}
 \caption{Comparison of the solutions for $\al$ and $y$ in a RG-invariant form for a different number of loops taken into account. Hereafter the plots are drawn not in terms of $L=\ln(Q^2/\mu^2)$ but rather $\hat L=\ln(Q^2/\Lambda^2)$\label{fig:compare_loops_np}}
\end{figure}

One can see that all curves merge at high energies but deviate closer to the pole where the perturbation theory fails.  At the same time, the more terms of expansion are taken into account the smoother is  the behaviour of the curve.

\section{Vertical Summation}

So far, we have obtained the following ``reduced'' form of our solutions:
\beqa
 \hat  \al_1 &= & \dfrac1{-\beta_0\hat L},  \\
  \hat \al_2 &= & \beta_0\hat \al_1^2\zeta_1, \label{vert1} \\
  \hat \al_3 &= & \beta_0^2\hat \al_1^3\zeta_1^2 + \hat \al_1^2 (\bt_{10} \hat \al_1 + \bt_{01} \hat y_1)\zeta_1 +  \beta_0\hat \al_1 ( c_5 \hat \al_1^2 + c_6 \hat \al_1\hat y_1 + c_7 \hat y_1^2 ),   \\
  \hat y_1 &= & \dfrac{\gm_{10}-\beta_0}{\beta_0\gm_{01}\hat L + K (\gm_{10}-\beta_0) \hat L^\nu}, 
   \\  \hat y_2& = & \hat y_1(\gm_{10}\hat \al_1+\gm_{01}\hat y_1)\zeta_1 + \hat y_1(c_3\hat \al_1+c_4\hat y_1),  \label{vert2}\\
  \hat y_3& = & \hat y_1 (\gm_{01} \hat y_1^2 + \dfrac32\gm_{10} \hat y_1\hat \al_1 + \dfrac{\gm_{10}(\gm_{10}+\beta_0)}{2\gm_{01}}\hat \al_1^2) \zeta_1^2  \nonumber \\
  &+&   \hat y_1 (c_8 \hat y_1^2 + c_9 \hat y_1\hat \al_1 + c_{10}\hat \al_1^2) \zeta_1 +  \hat y_1 (c_{11}\hat y_1^2 + c_{12}\hat y_1 \hat \al_1 + c_{13} \hat \al_1^2),  \\
  \dots  && \nonumber 
\eeqa
with 
$
\zeta_1=c_1 \ln\hat \al_1+c_2\ln\hat y_1.
$

We now try to sum the "vertical" infinite series of $\ln\hat \al_1$ and $\ln\hat y_1$.  We denote the resulting expressions as $\hat\al_\ell^{(1)}$ and $\hat y_\ell^{(1)}$. It turns out that they are given by the same functions with a replacement of arguments. 
It fact,  $\hat\al_\ell^{(1)}$ and $\hat y_\ell^{(1)}$ repeat the form of $\al_\ell^R$ and $y_\ell^R$ with the replacement:
\beqa
    \al_\ell^{(1)} &=& \al_\ell^R (L\to\zeta_1, \al_0\to\hat \al_1, y_0\to\hat y_1), \label{vert1} \\
    y  _\ell^{(1)} &=& y  _\ell^R (L\to\zeta_1, \al_0\to\alh_1, y_0\to\yh_1)  \label{vert2},
\eeqa
which can be checked directly by expanding $\al_\ell^R$ and $y_\ell^R$ in a series of $L,\al_0,y_0$. Alternatively, we can write equations (\ref{vert1},\ref{vert2}) in terms of $\alh_1,\yh_1$:
\begin{equation}\label{eq:al_y_beautiful_vertical_hats}
  \begin{array}{l}
    \hat \al_1^{(1)} = \dfrac{\hat \al_1}{1-\beta_0\hat \al_1\zeta_1},\\[10pt]
    \hat y _1^{(1)} = \dfrac{(\gm_{10}-\beta_0)\hat \al_1{\hat y_1}}{-\gm_{01}{\hat y_1} (1-\beta_0\hat \al_1\zeta_1) + (\hat \al_1 (\gm_{10}-\beta_0) + \gm_{01} \hat y_1)(1-\beta_0\hat \al_1\zeta_1)^{\scalebox{1.}{$\nu$}}} , \\[20pt]
    \hat \al_\ell^{(1)} = \hat \al_\ell\left[\hat \al_1\to\hat \al_1^{(1)},\,\hat y_1\to \hat y_1^{(1)},\,\ln\hat \al_1\to\ln\hat \al_1^{(1)}/\hat \al_1,\,\ln\hat y_1\to\ln \hat y_1^{(1)}/\hat y_1\right],\\[8pt]
    \hat y_\ell^{(1)} = \hat y_\ell\left[\hat \al_1\to\hat \al_1^{(1)},\,\hat y_1\to \hat y_1^{(1)},\,\ln\hat \al_1\to\ln\hat \al_1^{(1)}/\hat \al_1,\,\ln\hat y_1\to\ln \hat y_1^{(1)}/\hat y_1\right],
  \end{array}
\end{equation}
which can be formally rewritten  as
\begin{equation}\label{eq:al_y_beautiful_vertical_hats_new}
  \begin{array}{l}
   \hat  \al_1^{(1)} = \dfrac{\hat \al_1}{1-\beta_0\hat \al_1\zeta_1}, \\[10pt]
    \hat y_1^{(1)} = \dfrac{(\gm_{10}-\beta_0)\hat \al_1{\hat y_1}}{-\gm_{01}{\hat y_1} (1-\beta_0\hat \al_1\zeta_1) + (\hat \al_1 (\gm_{10}-\beta_0) + \gm_{01} \hat y_1)(1-\beta_0\hat \al_1\zeta_1)^{\scalebox{1.}{$\nu$}}}, \\[10pt]
   \hat  \al_\ell^{(1)} = \Xi_1^\ell\hat \al_\ell\left[\hat \al_1^{(1)}/\Xi_1,\, \hat y_1^{(1)}/\Xi_1\right],\\[8pt]
   \hat  y_\ell^{(1)} = \Xi_1^\ell\hat y_\ell\left[\hat \al_1^{(1)}/\Xi_1,\, \hat y_1^{(1)}/\Xi_1\right], \\[10pt]
    \Xi_1 = \left(\hat \al_1^{c_1}\hat y_1^{c_2}\right)^{\frac1{c_1+c_2}}.
  \end{array}
\end{equation}
One should not be confused by the appearance of the object $\Xi_1$ with non-integer powers of couplings, since they disappear when substituted inside the logarithms. Indeed, $\ln \Xi_1= \frac{c_1}{c_1+c_2} \ln \hat \al_1+\frac{c_2}{c_1+c_2} \ln\hat y_1$  and powers of $\Xi_1$ in (\ref{eq:al_y_beautiful_vertical_hats_new}) just cancel. This can be seen from the explicit  expressions in the lowest orders:
\begin{equation}\label{eq:al_y_vertical_explicit_2}
  \begin{array}{l}
   \hat  \al_2^{(1)} = \beta_0(\hat \al_1^{(1)})^2(c_1 \ln(\hat \al_1^{(1)}/\hat \al_1)+c_2\ln(\hat y_1^{(1)}/\hat y_1)),  \\[10pt]
    \hat y_2^{(1)} = \hat y_1^{(1)} (\gm_{10}\hat  \al_1^{(1)} + \gm_{01}  \hat y_1^{(1)}) (c_1 \ln(\hat \al_1^{(1)}/\hat \al_1)+c_2\ln(\hat y_1^{(1)}/\hat y_1)) + \hat y_1^{(1)} (c_3 \hat \al_1^{(1)} + c_4 \hat y_1^{(1)} ), \\[10pt]
    \hat \al_3^{(1)} = \beta_0^2(\hat \al_1^{(1)})^3 (c_1 \ln(\hat \al_1^{(1)}/\hat \al_1)+c_2\ln(\hat y_1^{(1)}/\hat y_1)^2 \\[8pt]\qquad + (\hat \al_1^{(1)})^2 (\bt_{10} \hat \al_1^{(1)} + \bt_{01} \hat y_1^{(1)}) (c_1 \ln(\hat \al_1^{(1)}/\hat \al_1)+c_2\ln(\hat y_1^{(1)}/\hat y_1)) \\[8pt]\qquad + \beta_0\hat \al_1^{(1)} ( c_5 (\hat \al_1^{(1)})^2 + c_6 \hat \al_1^{(1)}\hat y_1^{(1)} + c_7 (\hat y_1^{(1)})^2 ),\\[10pt]
    \hat y_3^{(1)}= \yh_1^{(1)} (\gm_{01} (\yh_1^{(1)})^2 + \frac32\gm_{10} \yh_1^{(1)}\alh_1^{(1)} + \frac{\gm_{10}(\gm_{10}+\beta_0)}{2\gm_{01}}(\alh_1^{(1)})^2) (c_1 \ln(\alh_1^{(1)}/\alh_1)+c_2\ln(\yh_1^{(1)}/\yh_1))^2 \\[10pt]\qquad + \yh_1^{(1)} \left(c_8 (\yh_1^{(1)})^2 + c_9 \yh_1^{(1)}\alh_1^{(1)} + c_{10}(\alh_1^{(1)})^2\right) (c_1 \ln(\alh_1^{(1)}/\alh_1)+c_2\ln(\yh_1^{(1)}/\yh_1)) \\[10pt]\qquad + \yh_1^{(1)} (c_{11}(\yh_1^{(1)})^2 + c_{12}\yh_1^{(1)} \alh_1^{(1)} + c_{13} (\alh_1^{(1)})^2).
  \end{array}
\end{equation}

Since the functions appearing at the level "1" are the same as the original solutions,  the logarithmic terms inside $\hat \al_\ell^{(1)}$ and $\hat y_\ell^{(1)}$ can once again be vertically summated. One has, respectively,
\begin{equation}\label{eq:al_y_beautiful_vertical_hats_2_new}
  \begin{array}{l}
   \hat  \al_1^{(2)} = \dfrac{\hat \al_1^{(1)}}{1-\beta_0\hat \al_1^{(1)}\zeta_2}, \\[10pt]
   \hat  y_1^{(2)} = \dfrac{(\gm_{10}-\beta_0)\hat \al_1^{(1)}{\hat y_1^{(1)}}}{-\gm_{01}{\hat y_1^{(1)}} (1-\beta\hat \al_1^{(1)}\zeta_2) + \left(\hat \al_1^{(1)} (\gm_{10}-\beta_0) + \gm_{01}\hat  y_1^{(1)}\right)(1-\beta_0\hat \al_1^{(1)}\zeta_2)^{\scalebox{1.}{$\nu$}}} , \\[10pt]
   \hat  \al_\ell^{(2)} = \Xi_2^\ell \hat \al_\ell\left[\hat \al_1^{(2)}/\Xi_2,\, \hat y_1^{(2)}/\Xi_2\right],\\[8pt]
    \hat y_\ell^{(2)} = \Xi_2^\ell \hat y_\ell\left[\hat \al_1^{(2)}/\Xi_2,\,\hat y_1^{(2)}/\Xi_2\right], \\[10pt]
  \zeta_2=c_1 \ln (\alh_1^{(1)}/\alh_1)+c_2\ln(\yh_1^{(1)}/\yh_1),  \ \ \  \Xi_2 = \left((\hat \al_1^{(1)})^{c_1}(\hat y_1^{(1)})^{c_2}\right)^{\frac1{c_1+c_2}}.
  \end{array}
\end{equation}

Moreover, we can perform this as many times as needed (here $n>2$):
\begin{equation}\label{eq:al_y_beautiful_vertical_hats_n_new}
  \begin{array}{l}
   \hat  \al_1^{(n)} = \dfrac{\hat \al_1^{(n-1)}}{1-\beta_0\hat \al_1^{(n-1)}\zeta_n}, \\[10pt]
   \hat  y_1^{(n)} = \dfrac{(\gm_{10}-\beta_0)\hat \al_1^{(n-1)}{\hat y_1^{(n-1)}}}{-\gm_{01}{\hat y_1^{(n-1)}} (1-\beta_0\hat \al_1^{(n-1)}\zeta_n) + (\hat \al_1^{(n-1)} (\gm_{10}-\beta_0) + \gm_{01} \hat y_1^{(n-1)})(1-\beta_0\hat \al_1^{(n-1)}\zeta_n)^{\scalebox{1.}{$\nu$}}},  \\[10pt]
    \hat \al_\ell^{(n)} = \Xi_n^\ell\hat \al_\ell\left[\hat \al_1^{(n)}/\Xi_n,\,\hat y_1^{(n)}/\Xi_n\right], \\[8pt]
   \hat  y_\ell^{(n)} = \Xi_n^\ell\hat y_\ell\left[\hat \al_1^{(n)}/\Xi_n,\,\hat  y_1^{(n)}/\Xi_n\right] ,\\[10pt]
  \zeta_n = c_1\ln(\hat \al_1^{(n-1)}/\hat \al_1^{(n-2)}) + c_2\ln(\hat y_1^{(n-1)}/\hat y_1^{(n-2)}),\ \ \   \Xi_n = \left((\hat \al_1^{(n-1)})^{c_1}(\hat y_1^{(n-1)})^{c_2}\right)^{\frac1{c_1+c_2}}.
  \end{array}
\end{equation}

One can notice the similarity between $\Xi_n$ and $\zeta_n$. Indeed, the connection between the two can be expressed as follows:
\begin{equation}\label{eq:zeta_n_Xi}
  \zeta_n = \left(c_1+c_2\right)\ln\left(\dfrac{\Xi_n}{\Xi_{n-1}}\right) \quad <=> \quad \Xi_n = \exp\left(\dfrac{\zeta_1+\ldots+\zeta_n}{c_1+c_2}\right).
\end{equation}
Note that if we go back to the case of the absence of $y$ in the beta-function for $\al$, we get $$\bt_{01}=0\,=>\,c_1 = \bt_{10}/\beta_0^2,\,c_2=0\,=>\,\Xi_n = \hat \al_1^{(n-1)}.$$

These formulas allow further improvement of the approximation due to further summation of the vertical infinite series, and this procedure continues endlessly. At each step, one has the same functions but with different arguments and one can cut this process at a point when a new term of the beta-function appears. For the calculated N-loop beta-functions one has  for the best approximation
\beq \label{prescription}
\bar \alpha=\sum_{\ell=1}^{N}\hat \alpha_\ell^{(N-1)},\ \  \bar y=\sum_{\ell=1}^{N}\hat y_\ell^{(N-1)},
\eeq
where $\hat \alpha_\ell^{(n)}$  and $\hat y_\ell^{(n)}$ are given by equations (\ref{eq:al_y_beautiful_vertical_hats_n_new}). 

\begin{figure}[ht]
 \centering
 \includegraphics[width=0.47\textwidth]{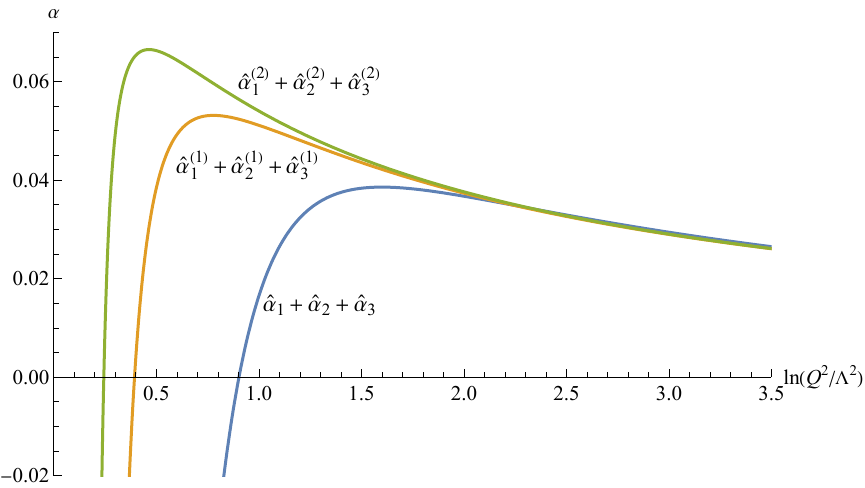}
 \includegraphics[width=0.47\textwidth]{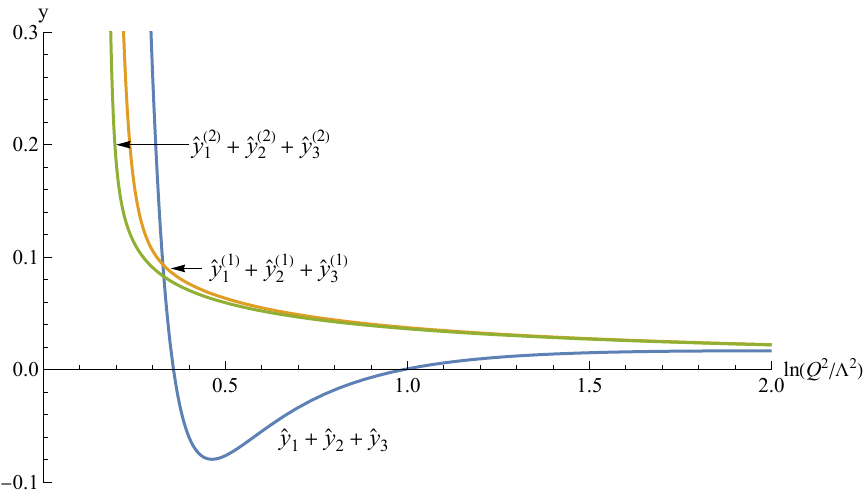}
 \caption{Solutions for $\al$ and $y$ in three-loop order at different stages of vertical resummation. Here one can see that the more times the vertical summation is performed, the further the solution extends, and the trustworthy interval approaches the pole.\label{fig:vertical_3}}
\end{figure}

We demonstrate in Fig.\ref{fig:vertical_3} the role of subsequent approximations when the three-loop beta-functions are taken into account. One can see that, while solutions without vertical summation (blue curves) are somewhat irregular near the IR pole, the behaviour of $\yh$ at 3 loops becomes monotonous  with one and two iterations of vertical summation included.

\begin{figure}[ht]
 \centering
 \includegraphics[width=0.45\textwidth]{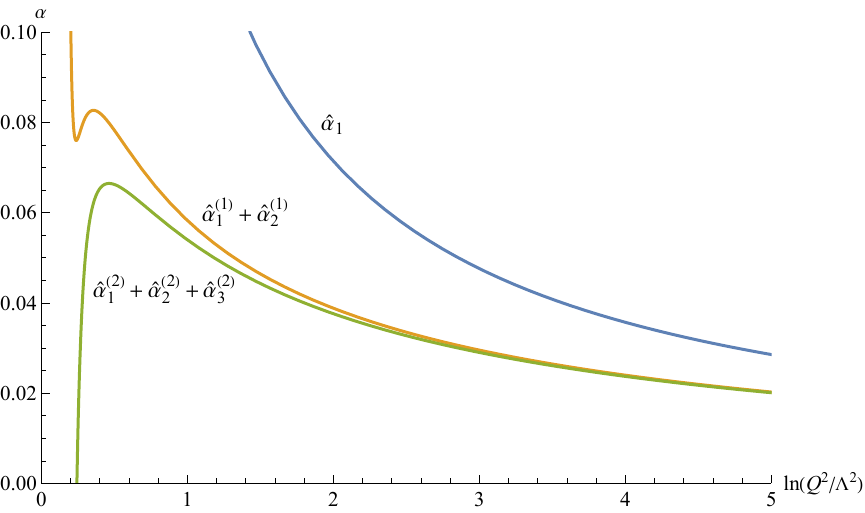}
 \includegraphics[width=0.45\textwidth]{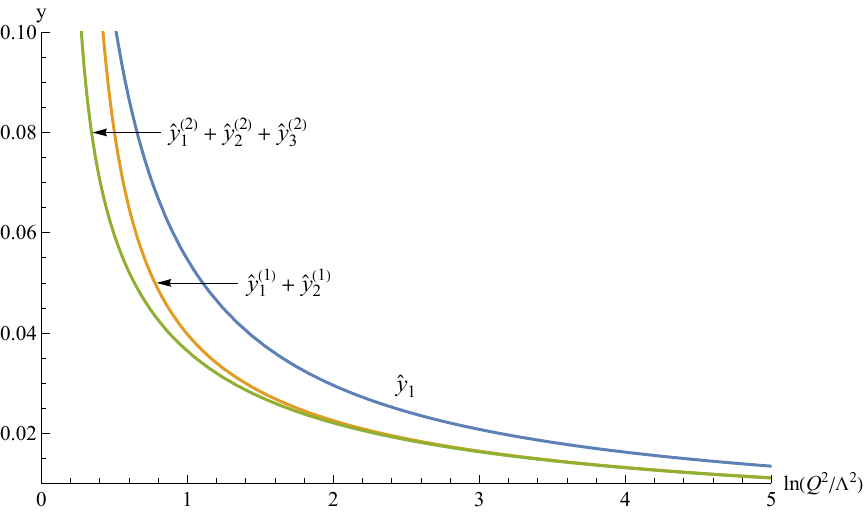}
 \caption{Comparison of the best possible approximations for $\al$ and $y$ in the cases of $1$, $2$ and $3$ loop orders of the beta-functions are available. \label{fig:cool_for_loop}}
\end{figure}

The next plot demonstrates how the addition of new loops  calculated according to the prescription (\ref{prescription}) improves the approximation. One can see that resummation indeed makes the behaviour of the solution more regular, approaching the smooth curve monotonously with a tendency to convergence of the procedure.

One can notice the unusual bending of $\alh$ near the pole in 2 loops. This is just a consequence of  the linearisation of equations.  Solving the equation for $\alpha_2$, we start with the initial condition $\alpha_2(0)=0$,
and as a result $\alpha_2$ becomes negative for a while before becoming positive again.  This does not influence  the total $\alpha=\alpha_1+\alpha_2$,  which stays always positive.
  
 \section{Conclusion}
 
 To conclude, we have demonstrated that choosing the right strategy, one can get relatively simple explicit formulas for the running coupling expansion in the asymptotic regime. These formulas contain only logarithms and no special functions, and correspond to the summation of infinite series of leading, next-to-leading, etc logarithms.  
 The case of two couplings appeared to be more complicated, and one has to neglect certain terms that do not allow a description in terms of elementary functions. However, in the asymptotic regime that  we consider, these terms give an inessential contribution. The strategy allows further improvement of the approximation by summing the "vertical" series of terms, which are described by the same functions with a change of  arguments. 
 
\section*{Appendix A: Coefficients $c_i$}
We present here the expressions for the coefficients $c_i$:
\begin{equation*}
\begin{array}{l}
  c_1 = \dfrac{\bt_{10} \gm_{01} - \bt_{01} \gm_{10}}{\beta_0^2\gm_{01}} \stackrel{SM}{=} -\dfrac{38}{63}\\[8pt]
  c_2 = \dfrac{\bt_{01}}{\beta_0\gm_{01}} \stackrel{SM}{=} \dfrac{4}{63} \\[8pt]
  c_3 = \dfrac{\beta_0  \gm_{20}-\bt_{10} \gm_{10}}{\beta_0 ^2} \stackrel{SM}{=} \dfrac{548}{49} \\[8pt]
  c_4 = \dfrac{-\beta_0 \bt_{01} \gm_{10}-\beta_0  \bt_{10} \gm_{01}-\bt_{10} \gm_{01} \gm_{10}+\beta_0  \gm_{01} \gm_{20}+\beta_0 ^2 \gm_{11}}{\beta_0 ^2 \gm_{10}} \stackrel{SM}{=} -\dfrac{3523}{392} \\[8pt]
  c_5 = \dfrac{\beta_0  \bt_{20}-\bt_{10}^2}{\beta_0 ^2} \stackrel{SM}{=} -\dfrac{1807}{98} \\[8pt]
  c_6 =  \dfrac{-2 \beta_0  \bt_{01} \bt_{10}-\bt_{01} \bt_{10} \gm_{10}+\beta_0  \bt_{01} \gm_{20}+\beta_0 ^2 \bt_{11}}{\beta_0^2 \gm_{10}} \stackrel{SM}{=} \dfrac{291}{49} \\[8pt]
  c_7 =  \dfrac{2 \bt_{01}^2 \gm_{10}-\bt_{01} \bt_{10} \gm_{01}-\beta_0  \bt_{01} \gm_{11}-\beta_0  \bt_{02} \gm_{10}+\beta_0  \bt_{11} \gm_{01}}{\beta_0  \gm_{10} (\beta_0 -2 \gm_{10})} \stackrel{SM}{=} -\dfrac{191}{252}\\[8pt]
  c_8 = \dfrac{\gm_{01} \left(-\beta_0  (\bt_{01} \gm_{10}+2 \bt_{10} \gm_{01}-2 \gm_{01} \gm{20})-2 \bt_{10} \gm_{01} \gm_{10}+2 \beta_0 ^2 \gm_{11}\right)}{\beta_0^2 \gm_{10}} \stackrel{SM}{=} -\dfrac{31203}{392}\\[8pt]
  c_9 = -\dfrac{\bt_{01} \gm_{10}+\bt_{10} \gm_{01}-3 \gm_{01} \gm_{20}}{\beta_0 }-\dfrac{3 \bt_{10} \gm_{01} \gm_{10}}{\beta_0^2}+2 \gm_{11} \stackrel{SM}{=} \dfrac{10219}{49} \\[8pt]
  c_{10} =  \dfrac{\beta_0  \gm_{20} (\beta_0 +\gm_{10})-\bt_{10} \gm_{10}^2}{\beta_0^2} \stackrel{SM}{=} -\dfrac{9676}{49}\\[8pt]
 \end{array}
 \end{equation*}
 \begin{equation*}
\begin{array}{l}
   c_{11} = \dfrac{\beta _0^5 (2 \gm_{11}^2\!+\!\gm_{10} \gm_{12})\!+\!\beta_0^4 (-\bt_{02} \gm_{10}^2\!+\!\bt_{11} \gm_{01} \gm_{10}\!-\!5 \bt_{01} \gm_{11} \gm_{10}\!-\!4 \bt_{10} \gm_{01} \gm_{11}\!-\!2 \gm_{12} \gm_{10}^2)}{2 \beta_0^4 \gm_{10}^2 (\beta_0\! -\!2 \gm_{10})}\\[8pt] 
   +\dfrac{ \beta_0^4(\!-\!4 \gm_{11}^2 \gm_{10}\!+\!3 \gm_{02} \gm_{20} \gm_{10} \!+\!4 \gm_{01} \gm_{11} \gm_{20})\!+\!\beta_0^3 (2 \bt_{10}^2 \gm_{01}^2\!+\!3 \bt_{01}^2 \gm_{10}^2\!+\!\bt_{01} \gm_{10} (8 \gm_{10} \gm_{11}\!-\!5 \gm_{01} \gm_{20}))}{2 \beta_0^4 \gm_{10}^2 (\beta_0 -2 \gm_{10})} \\[8pt]
   +\dfrac{ \beta_0^3 (\bt_{10} (\bt_{01} \gm_{10} \gm_{01}\!-\!4 \gm_{20} \gm_{01}^2 \!+\! 4 \gm_{10} \gm_{11} \gm_{01} )\! -\!3 \bt_{10}\gm_{02} \gm_{10}^2\!+\!2 \gm_{20} (\gm_{20} \gm_{01}^2\!-\!4 \gm_{10} \gm_{11} \gm_{01}\!-\!3 \gm_{02} \gm_{10}^2))}{2 \beta_0^4 \gm_{10}^2 (\beta_0 -2 \gm_{10})} \\[8pt]
   +\dfrac{ \beta_0^2 (\gm_{10} \!-\!2 \bt_{01}^2 \gm_{10}^2\!+\!\bt_{01} \gm_{01} \gm_{10} (\bt_{10}\!+\!10 \gm_{20})\!+\!\gm_{10} (\bt_{10} (4 \gm_{20} \gm_{01}^2\!+\!8 \gm_{10} \gm_{11} \gm_{01}\!+\!6 \gm_{02} \gm_{10}^2)\!-\!4 \gm_{01}^2 \gm_{20}^2)}{2 \beta_0^4 \gm_{10}^2 (\beta_0 -2 \gm_{10})} \\[8pt]   
   -\dfrac{ -2 \beta_0  \bt_{10} \gm_{01} \gm_{10}^2 (3 \bt_{10} \gm_{01}\!+\!5 \bt_{01} \gm_{10}\!-\!4 \gm_{20} \gm_{01})\!+\!4 \bt_{10}^2 \gm_{01}^2 \gm_{10}^3}{2 \beta_0^4 \gm_{10}^2 (\beta_0 -2 \gm_{10})} \stackrel{SM}{=}\dfrac{161562269}{1382976} \\
    c_{12} =  \dfrac{2 \beta_0^4 (\bt_{11} \gm_{10}\!+\!\gm_{21} \gm_{10}\!+\!\gm_{11} \gm_{20})\!+\!\beta_0^3 (-2 \bt_{01} \gm_{10} (2 \bt_{10}\!+\!\gm_{20})\!-\!2 \bt_{10} (2 \gm_{10} \gm_{11}\!+\!\gm_{01} \gm_{20}))}{2 \beta_0^4 \gm_{10} (\gm_{10}+\beta_0 )} \\[8pt]
   +\dfrac{ \beta_0^3 (2 \gm_{01} \gm_{20}^2\!+\!4 \gm_{10} \gm_{11} \gm_{20}\!+\!\gm_{01} \gm_{10} \gm_{30})\!+\!\beta_0^2 \gm_{10} (2 \bt_{10}^2 \gm_{01}\!-\!\bt_{10} (4 \gm_{10} \gm_{11}\!+\!9 \gm_{01} \gm_{20})\!+\!\bt_{20} \gm_{01} \gm_{10})}{2 \beta_0^4 \gm_{10} (\gm_{10}+\beta_0 )} \\[8pt]
   +\dfrac{\bt_0^2\gm_{20} (3 \gm_{01} \gm_{20}-2 \bt_{01} \gm_{10})+\beta_0  \bt_{10} \gm_{10}^2 (5 \bt_{10} \gm_{01}+2 \bt_{01} \gm_{10}-6 \gm_{20} \gm_{01})+3 \bt_{10}^2 \gm_{01} \gm_{10}^3}{2 \beta_0^4 \gm_{10} (\gm_{10}+\beta_0 )} \\[8pt]
   \stackrel{SM}{=} \dfrac{1012 \zeta (3)}{35}-\dfrac{39792677}{216090} \\[8pt]
  c_{13} =   \dfrac{\bt_{10}^2 \gm_{10}^2-\beta_0  \bt_{10}^2 \gm_{10}-2 \beta_0  \bt_{10} \gm_{10} \gm_{20}-\beta_0^2 \bt_{10} \gm_{20}+\beta_0^2 \bt_{20} \gm_{10}+\beta_0^2 \gm_{20}^2+\beta_0^3 \gm_{30}}{2 \beta_0^4} \\[8pt]
  \stackrel{SM}{=} \dfrac{882643}{7203}-\dfrac{320 \zeta (3)}{7} \\[8pt]
  c_{14} = \dfrac{\beta_0 (\bt_{11} \gm_{10}+\gm_{11} \gm_{20})-\bt_{10} \gm_{01} \gm_{20}-2 \bt_{01} \bt_{10} \gm_{10}+\bt_{20} \gm_{01} \gm_{10}-\bt_{10} \gm_{10} \gm_{11}+\gm_{01} \gm_{20}^2}{\beta_0^2 \gm_{10}}  \\[8pt] +  \dfrac{-2 \beta_0  \bt_{10} \gm_{01} \gm_{10} \gm_{20}+\bt_{10}^2 \gm_{01} \gm_{10}^2}{\beta_0^4 \gm_{10}} \stackrel{SM}{=} -\dfrac{786277}{9604}
\end{array}
\end{equation*}

 \section*{Acknowledgments}
 The authors thank A.Bednyakov and S.Mikhailov for useful comments and discussions. A.F. is grateful to the Laboratory of Theoretical Physics for hospitality.

 \end{document}